%% file: RMmetal-jan01.tex
\documentclass[useAMS,usenatbib]{mn2e}
\include{epsf}
\include{journals}
\usepackage{amsmath}
\usepackage{graphicx}
\usepackage{subfig}
\usepackage{float}
\usepackage{upquote}
\usepackage{footnote}
\usepackage{auto-pst-pdf}
\usepackage{epstopdf}
\DeclareGraphicsExtensions{.ps}
\input{aas_macros}

\title[Radial migration and the radial metallicity distribution]{Impact of radial migration on stellar and gas radial metallicity distribution}
\author[Grand et al.]
{Robert J.J. Grand$^{123}$\thanks{robert.grand@h-its.org}, Daisuke Kawata$^3\thanks{d.kawata@ucl.ac.uk}$ and Mark Cropper$^3$\\
$^1$Heidelberger Institut f{\"u}r Theoretische Studien, Schloss-Wolfsbrunnenweg 35, 69118 Heidelberg, Germany\\
$^2$Zentrum f{\"u}r Astronomie der Universit{\"a}t Heidelberg, Astronomisches Recheninstitut, M{\"o}nchhofstr. 12-14, 69120 Heidelberg, Germany \\
$^3$Mullard Space Science Laboratory, University College London, Holmbury St. Mary, Dorking, Surrey, RH5 6NT, United Kingdom}

\pagerange{\pageref{firstpage}--\pageref{lastpage}} \pubyear{2012}

\pagerange{\pageref{firstpage}--\pageref{lastpage}}\pubyear{2011}
\def\LaTeX{L\kern-.36em\raise.3ex\hbox{a}\kern-.15em
  T\kern-.1667em\lower.7ex\hbox{E}\kern-.125emX}
    
\begin{document}

\label{firstpage}
\maketitle

\begin{abstract}
Radial migration is defined as the change in guiding centre radius of stars and gas caused by gains or losses of angular momentum that result from gravitational interaction with non-axisymmetric structure. This has been shown to have significant impact on the metallicity distribution in galactic discs, and therefore affects the interpretation of Galactic archeology. We use a simulation of a Milky Way-sized galaxy to examine the effect of radial migration on the star and gas radial metallicity distribution. We find that both the star and gas component show significant radial migration. The stellar radial metallicity gradient remains almost unchanged but the radial metallicity distribution of the stars is broadened to produce a greater dispersion at all radii. However, the metallicity dispersion of the gas remains narrow. We find that the main drivers of the gas metallicity distribution evolution are metal enrichment and mixing: more efficient metal enrichment in the inner region maintains a negative slope in the radial metallicity distribution, and the metal mixing ensures the tight relationship of the gas metallicity with the radius. The metallicity distribution function reproduces the trend in the age-metallicity relation found from observations for stars younger than 1.0 Gyr in the Milky Way.
\end{abstract}


\section{Introduction}

The aim of Galactic archeology is to determine the formation history and evolution of the Milky Way by studying the metallicity and kinematic information of individual stars \citep[see][for a review]{RB13R}. This has spurred a number of Galactic surveys aimed at collecting this information for large samples of stars. Recent surveys include SEGUE \citep{YRN09,LBA11}, RAVE \citep{SZ06} and APOGEE \citep{APM08}, which together will obtain chemo-kinematical information for hundreds of thousands of stars within a few kiloparsecs of the solar neighbourhood. In the near future, projects such as 4MOST \citep{dJ12}, LAMOST \citep{CHY12}, $Gaia$-ESO \citep{GRAB12} and $Gaia$ \citep{LBB08} will provide chemo-kinematical data of millions of disc stars that cover a large portion of the Galactic disc.

The principle behind Galactic archeology is that the chemical abundances of a star are indicative of its age and origin. Chemical elements that are synthesised inside stars are distributed and mixed into the interstellar medium in the form of supernova ejecta. New generations of stars that form out of the chemically evolved gas effectively lock up these elements. Therefore the abundances of metals depend on the amount of preceding star formation, which is in turn tied to the age and location of birth of stars. This valuable chemical tag \citep{FBH02} can be combined with kinematic information of stars in order to identify the origin of different stellar populations and constrain the assembly history of the Milky Way. 

Complications for Galactic archeology arise especially in systems with non-axisymmetric structures such as stellar bars and spiral arms, which cause stars to move away from their birth radii through the process known as radial migration \citep{LBK72}. \citet{SB02} showed that radial migration can occur around the co-rotation radii of spiral arms where stars can gain and lose angular momentum without changing their random energy component. Because the stars rotate at the same speed as the spiral arm at co-rotation, the stars feel a continuous gravitational torque from the spiral arm, which leads to significant changes in the angular momentum, and therefore guiding centre, of the stars. In recent years, several variations of the radial migration mechanism have been proposed, such as bar induced migration \citep[e.g.][]{BCP11,SoS12}, multiple bar/spiral modes \citep[e.g.][]{MQ06,R08,RD11,MFQD12,CQ12}, satellite induced migration \citep{BiKW12} and co-rotating spiral arms that cause radial migration at every radius along the spiral arm \citep{GKC11,GKC12,BSW12,GKC13b}. In the last mechanism, the rotation speed of the spiral arm follows that of the disc mean rotation velocity, therefore stars feel a continuous gravitational force from the spiral arm at all radii during their migration. This can produce significant changes in the guiding centres of individual stars and hence streaming motions along the spiral arms, the features of which may be detectable in phase space \citep{KHG14}.

Radial migration has been highlighted in several studies as causing significant and lasting changes in the metallicity distribution in the disc \citep[][]{SB02,LA03,ScB09}. In particular, radial migration may influence the evolution of the radial metallicity gradient and cause an increased scatter in the metallicity distribution function \citep[e.g.][]{CM03,AG09,SBl09,PFG12,RKA11,RCK13,DiM13,KPA13,MCM13,SBl14,MCM14}. The increased scatter is seen in observations collected from the Geneva Copenhagen Survey \citep{NMA04,HNA09}, which revealed that the metallicity distribution function in the solar neighbourhood is broader for older stars than for the younger stars \citep{H08,CSA11}. Hence, as a result of radial migration, a large scatter is introduced in the age-metallicity relation at a given radius, and it also introduces additional uncertainty in galaxy assembly scenarios such as the formation of thick discs \citep{ANS03,Br04,VH08,BEM09,BSG12,Min12,Rok12,BRH12,BM13,VC14}.

In addition to the stellar component, gas is also subject to the gravity of the spiral arms and therefore to radial migration. Because the interstellar medium determines the metallicity distribution of young and newborn stars, the evolution of the gas component plays an important role in the overall evolution of the disc. However, there are few studies of the chemo-dynamical evolution of the gas component with respect to radial migration. \citet{MCM14} modelled the chemical evolution of a Milky Way-like simulated disc by combining cosmological numerical simulations with a semi-analytic chemical evolution model. They found that the radial migration of gas had little effect on the radial metallicity gradient. 

In this paper, we aim to complement and build upon previous studies. We examine the evolution of both star and gas particles in a simulation of a Milky Way-sized galaxy in an isolated setting with self-consistent chemo-dynamical evolution. We focus on the effects of radial migration coupled with the chemical evolution on the metallicity distribution and the radial metallicity gradient of the stars and gas. 

This paper is organised as follows. In section 2, we describe the numerical code and the simulated galaxy. In section 3, we describe the evolution of the star and gas radial metallicity distribution and the metallicity distribution function, and discuss the results. In section 4, we summarise our findings.

\section{Simulation}

\subsection{Numerical simulation code}

We use a $N$-body smoothed particle hydrodynamics (SPH) code, GCD+, which simulates galaxy formation and galactic chemo-dynamical evolution \citep[][]{KG03,RK11,BKW12,KOG13,KGBG13}. The code is based on the SPH method described in \citet{L77,GM77}, and includes self-gravity, hydrodynamics, radiative cooling, star formation, supernovae (SNe) feedback and metal enrichment and diffusion \citep{GGB09}. The hydrodynamics calculation employs the adaptive softening length scheme suggested by \citet{PM07}. GCD+ uses the artificial viscosity switch used by \citet{MM97} and artificial thermal conductivity in order to resolve the Kelvin-Helmholtz instability \citep[see][]{P08,RHA10,H13b,SM13}. The code integrates the entropy equation \citep{SH02} as well as the hydrodynamical equations. The code adopts the individual time-step limiter \citep{SM09} which is necessary in order to correctly resolve the expansion of SNe-blown bubbles. The FAST integration scheme \citep{SM10} is implemented, which increases the calculation speed of the equations of motion by using different time-steps to integrate hydrodynamics and gravity. 

Radiative cooling and heating are calculated using CLOUDY \citep{FKV98,RK08}. The cooling and heating rates and the mean molecular weight are tabulated as a function of metallicity, density and temperature using the \citet{HM96} UV background radiation. A gas density threshold is employed to control the star formation rate: gas particles that exceed the threshold and are in a region of convergent velocity may change to star particles according to a probability weighted by the particle density. The details of the star formation and feedback recipe are fully described in \citet{KGBG13}. 

Each star particle in the simulation is assigned to a mass group according to a Salpeter initial mass function \citep{S55}. As described in \citet{RK11} and \citet{KGBG13}, star particles may (depending on the age of the star particle) change to feedback particles, which are responsible for the metal enrichment from supernova Type II and Ia tabulated from \citet{WW95} and \citet{I99} respectively. The metals from supernovae are deposited in only the feedback particles.

In addition to metal enrichment of feedback particles, metal diffusion between gas particles is modelled following the scheme outlined by \citet{GGB09}. In this scheme, the rate of change of the mass of each metal, $m_{Z,i}$, is described by the diffusion equation,

\begin{equation}
\frac{dm_Z}{dt} = \frac{1}{\rho}\nabla \cdot (D\nabla m_Z).
\label{eqdiffusion}
\end{equation}
In the SPH scheme, the evolution of the metallicity for the $i$-th particle is calculated from the discrete summation over all neighbour particles within the smoothing length,

\begin{equation}
\frac{dm_{Z,i}}{dt} = \sum_j K_{ij} (m_{Z,i} - m_{Z,j}), 
\label{eqdif}
\end{equation}
where

\begin{equation}
K_{ij} = \frac{m_j}{\rho _i \rho _j} \frac{4 D_i D_j}{(D_i + D_j)} \frac{(\mathbf{r}_i-\mathbf{r}_j) \cdot \nabla _i W_{ij}}{|\mathbf{r}_i-\mathbf{r}_j|^2},
\end{equation}
where $m$ is the particle mass, $\rho$ is the density, $\mathbf{r}_i-\mathbf{r}_j$ is the vector distance between the $i$-th and $j$-th particles and $W_{ij}$ is the SPH kernel. The diffusion coefficient of the $i$-th particle is given by

\begin{equation}
D_i = \rho _i h_i \sqrt{\frac{1}{N_{\rm nb}} \sum_j |\mathbf{v}_i - \mathbf{v}_j|^2},
\label{eqdc}
\end{equation}
where $h_i$ is the smoothing length, $N_{\rm nb}$ is the number of neighbour particles within $h_i$ and the square root term is the velocity dispersion evaluated for the $i$-th particle. Equation (\ref{eqdif}) is evaluated for every gas particle in the simulation, and is numerically integrated over the same time-steps as the dynamical time-step in the simulation. 

\subsection{Simulation}
\label{sec2}

The simulated Milky Way-sized galaxy analysed in this paper is the same as that presented in \citet{KHG14}. The galaxy is set up in isolated conditions, and consists of a gas and stellar disc with no bulge component. The discs are embedded in a static dark matter halo potential \citep{RK11,Hun13,KHG14}. The total mass of the dark matter halo is $M_{\rm dm}=2.5 \times 10^{12}$ $\rm M_{\odot}$, and the dark matter density follows the Navarro-Frenk-White (NFW) density profile \citep{NFW97}, with a concentration parameter of $c=10$. The stellar disc is assumed to follow an exponential surface density profile with the initial mass of $M_{\rm d,*} = 4.0 \times 10^{10}$ $\rm M_{\odot}$, a radial scale length of $R_{\rm d,*} = 2.5$ kpc and a scale height of $z_{\rm d,*} = 350$ pc. The gas disc is set up following the method of \citet{Sp05}, and follows an exponential surface density profile with a scale length, $R_{d,g} = 8.0$ kpc. The total gas mass is $1.0 \times 10^{10}$ $\rm M_{\odot}$. The simulation comprises $N=1 \times 10^6$ gas particles and $4 \times 10^6$ star particles; therefore each particle has a mass of $10, 000$ $\rm M_{\odot}$. The resolution is sufficient to minimise numerical heating from Poisson noise \citep{Fu11,Se13}. We apply a minimum softening length of $158$ pc (Plummer equivalent softening length of $53$ pc) with the spline softening and variable softening length for gas particles suggested by \citet{PM07}. 

The initial radial profile of the mean metallicity of stars and gas is set by

\begin{equation}
\mathrm{[Fe/H]} (R) = 0.2 - 0.05 \left( \frac{R}{1 \mathrm{kpc}} \right).
\end{equation}
The metallicity distribution function at each radius is centred on the mean metallicity value with the dispersion set to a Gaussian distribution of $0.05$ dex for the gas and $0.2$ dex for the stars. 

\begin{figure}
\includegraphics[scale=1.9] {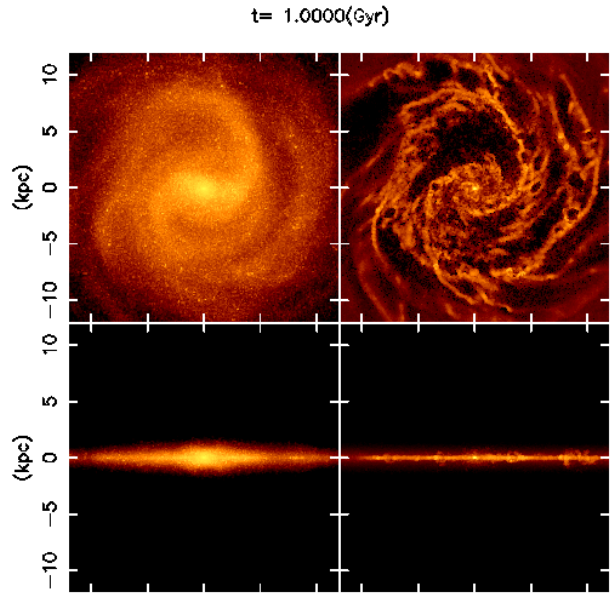} 
\caption{A snapshot of the star (left) and gas (right) distribution at $t=1.0$ Gyr. The upper (lower) panels show the face-on (side-on) view respectively.}
\label{figsnap}
\end{figure}

\section{Results}

\begin{figure*}
\begin{tabular} {l r} 
\includegraphics[scale=0.4] {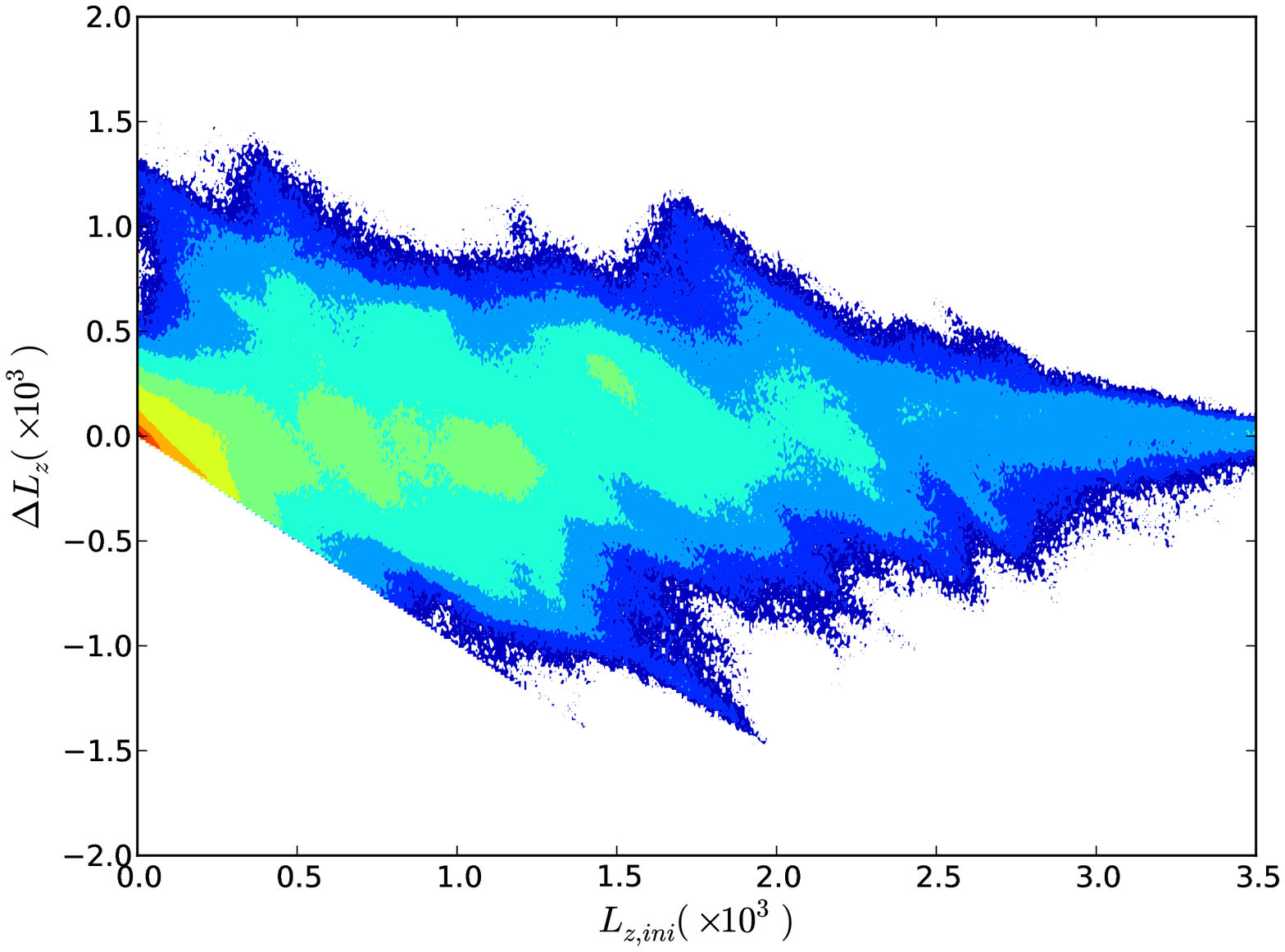} &  \includegraphics[scale=0.4] {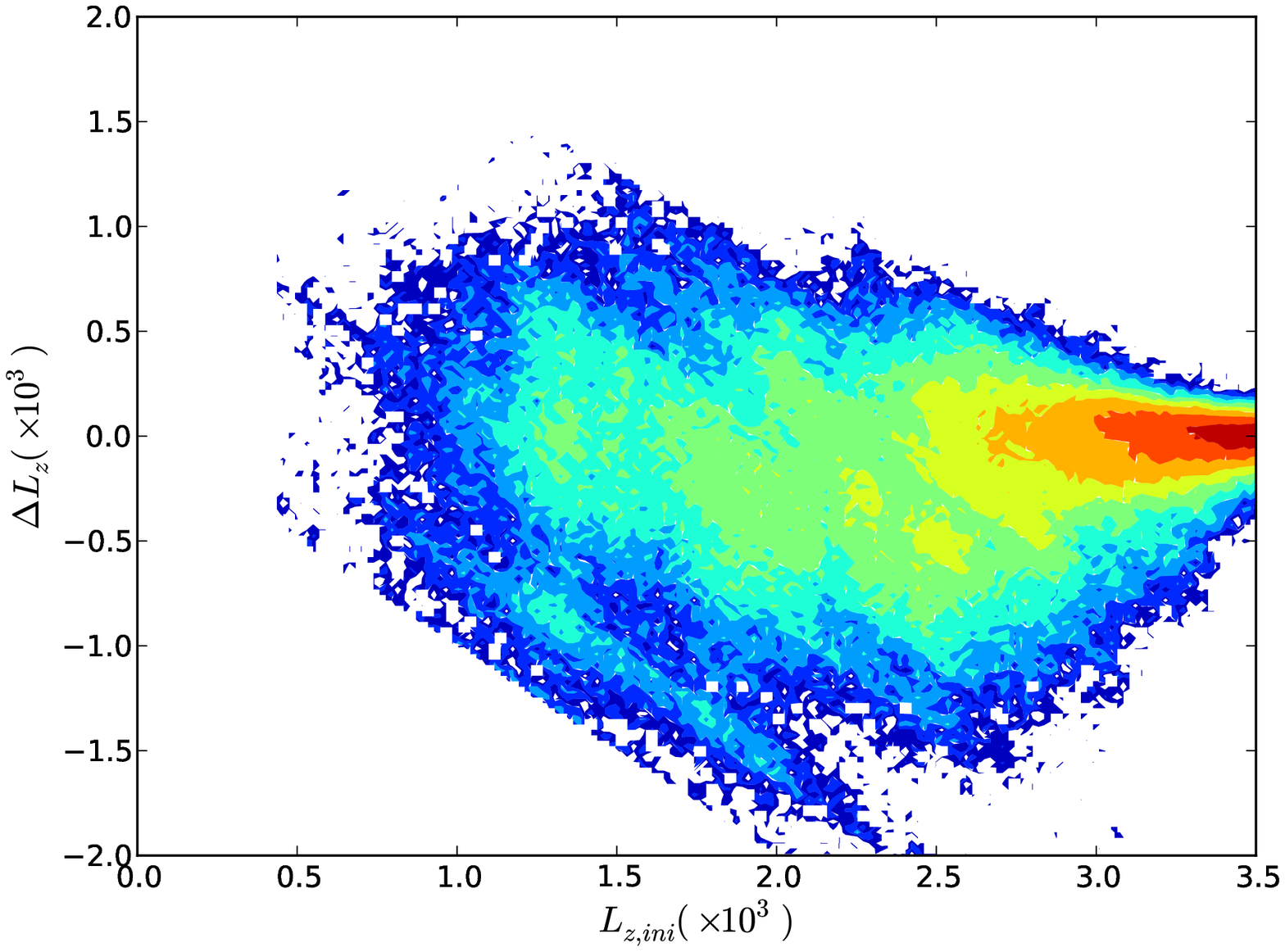} \\
\end{tabular}
\caption{The angular momentum change of all star particles (left) and all gas particles (right) between $t=0.5$ and $1.0$ Gyr, as a function of their angular momentum value at $t=0.5$ Gyr.}
\label{fig1}
\end{figure*}

\begin{figure*}
\includegraphics[scale=0.15] {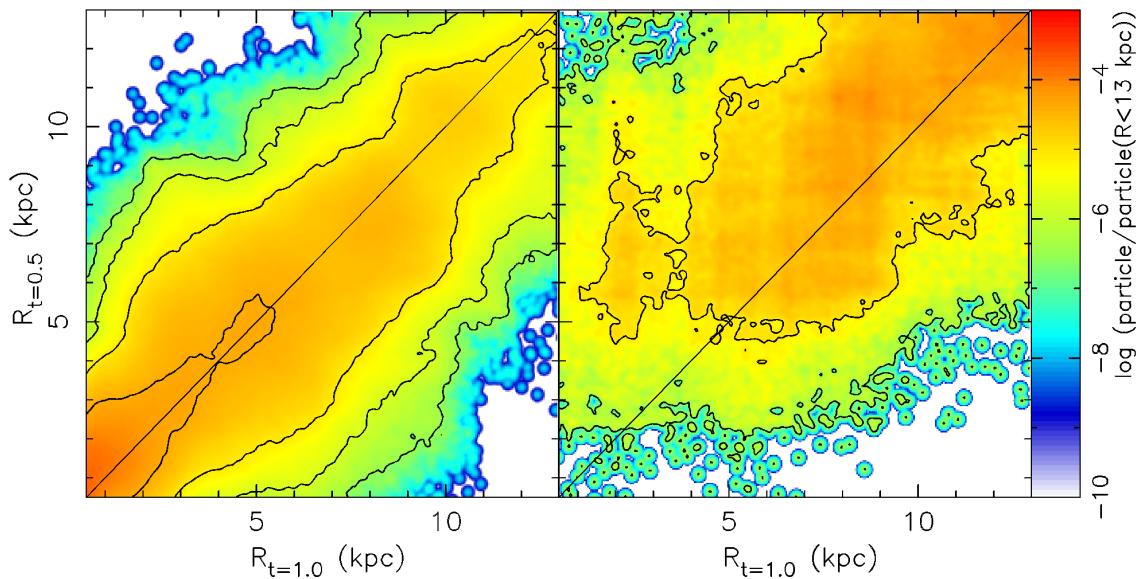} 
\caption{The radii at $t=0.5$ of all star particles (left) and all gas particles that did not turn into star particles between $t=0.5$ and $1.0$ Gyr (right) plotted as a function of their radii at $1.0$ Gyr. The colours represent the particle number normalised to the total star and gas particle number within $R=13$ kpc, respectively. The solid line indicates the $R_{t=0.5} = R_{t=1.0}$ relation.}
\label{figr}
\end{figure*}

The focus of this paper is the chemo-dynamical evolution of the gas and stellar disc, specifically the effects of radial migration on the metal distribution. In order to analyse these effects clearly, we focus on the period of evolution between $t=0.5$ and 1.0 Gyr. This period comes after much of the vigorous bar formation that takes place in the first 0.5 Gyr of the evolution, during which mass re-distribution leads to changes in the angular momentum profile of the system. By avoiding this unstable period, we are able to minimise the influence of angular momentum redistribution on the disc evolution, and focus on the effect of radial migration from the non-axisymmetric structures. We follow the evolution to 1.0 Gyr, because the consumption of gas from star formation and our simple numerical setting of no gas infall leads to very little star formation after 1.0 Gyr. Fig. \ref{figsnap} shows a snapshot of the simulated galaxy at $t=1.0$ Gyr.

\subsection{Radial migration of stars and gas}

The bar and spiral arms have been shown in many previous studies to induce angular momentum changes of star particles \citep[see][for recent reviews]{S14R,DB14}. The left panel of Fig. \ref{fig1} shows the angular momentum change, $\Delta L_z$, of all the star particles between the $0.5$ Gyr and $1.0$ Gyr time-steps as a function of their angular momentum, $L_{z,ini}$, at $0.5$ Gyr. The broad swathes of non-zero $\Delta L_z$ that cover all angular momentum values (therefore, all radii) indicate that radial migration occurs at all radii of the disc. The significant radial migration that occurs over short time periods has been reported in our previous studies in which the spiral arms were found to co-rotate with the star particles \citep{GKC11,GKC12,GKC13b}. Therefore although we follow here only a relatively short period of evolution, this is still sufficient to study the effect of radial migration.

The right panel of Fig. \ref{fig1} shows the same as the left panel for the gas particles. A broad distribution similar to that of the star particles is seen, which confirms that the gas particles also radially migrate at all radii as a consequence of their interaction with the spiral arms and bar. The figure plots only those gas particles that existed at both $t=0.5$ and 1.0 Gyr: gas particles that existed at $t=0.5$ Gyr and turned into star particles before $t=1.0$ Gyr are not plotted. The paucity of gas particles at low values of $L_{z,ini}$ is caused by the more rapid star formation that follows from the high gas density in the central region of the disc; therefore very few low angular momentum gas particles survive until $t=1.0$ Gyr. This reasoning explains also the high density of gas particles at large values of $L_{z,ini}$.

\begin{figure}
\includegraphics[scale=0.9] {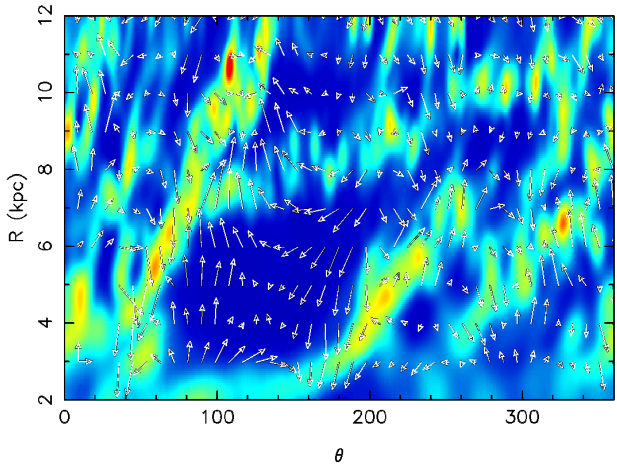} 
\caption{Smoothed density map of the gas in 2D polar coordinates at $t=0.7$ Gyr. The rotation direction is from right to left. The mean velocity field of the gas is shown by the arrows. The arrows in the bottom-left corner of the figure show the 20 km s$^{-1}$ vectors in both coordinates.}
\label{rthmap}
\end{figure}

The changes in radius for star and gas particles between $t=0.5$ and 1.0 Gyr are quantified in Fig. \ref{figr}, which shows the ``initial'' radius at $t=0.5$ Gyr plotted against the  ``final'' radius at $t=1.0$ Gyr for all star particles (left panel) and the gas particles that did not change to star particles within this period (right panel). The wide-spread distributions around the one to one relation (solid line) highlight large changes in particle radius over time that cannot be explained by epicycle motion alone \citep[e.g.][]{RDS08}. The degree of spread in the initial-final radius distribution of the gas particles (right panel) is similar to that of the star particles (left panel), though the dearth of gas particles at lower initial radii (or lower angular momentum as shown in Fig. \ref{fig1}) reveals an asymmetry in the diagram that is not present for the stars.

\begin{figure*}
\includegraphics[scale=0.16]{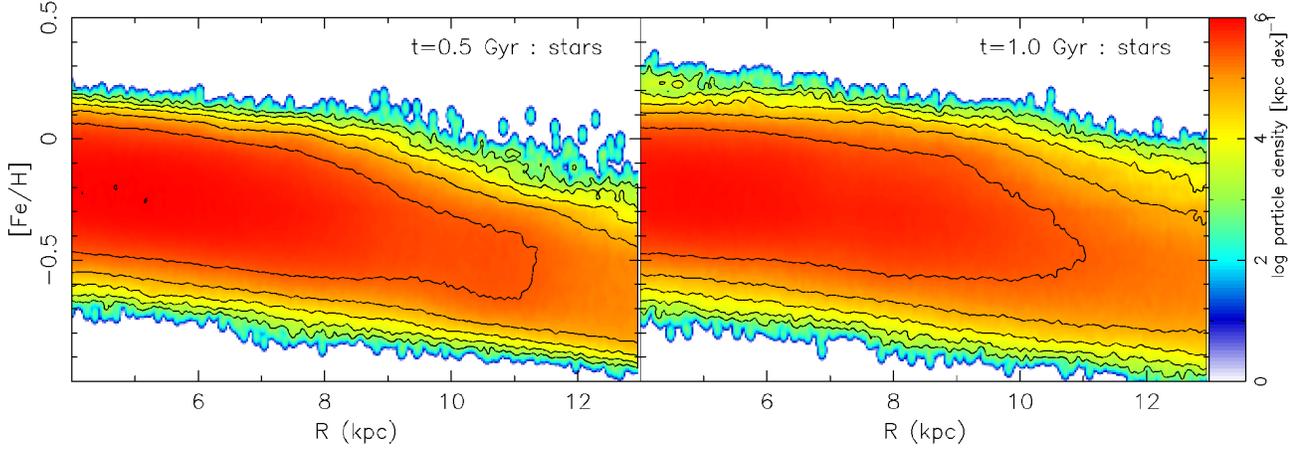} 
\caption{The RMD of all stars at $t=0.5$ Gyr (left) and all stars at $t=1.0$ Gyr (right). The redder colours indicate regions of higher number density.}
\label{fig2a}
\end{figure*}

\begin{figure*}
\includegraphics[scale=0.16]{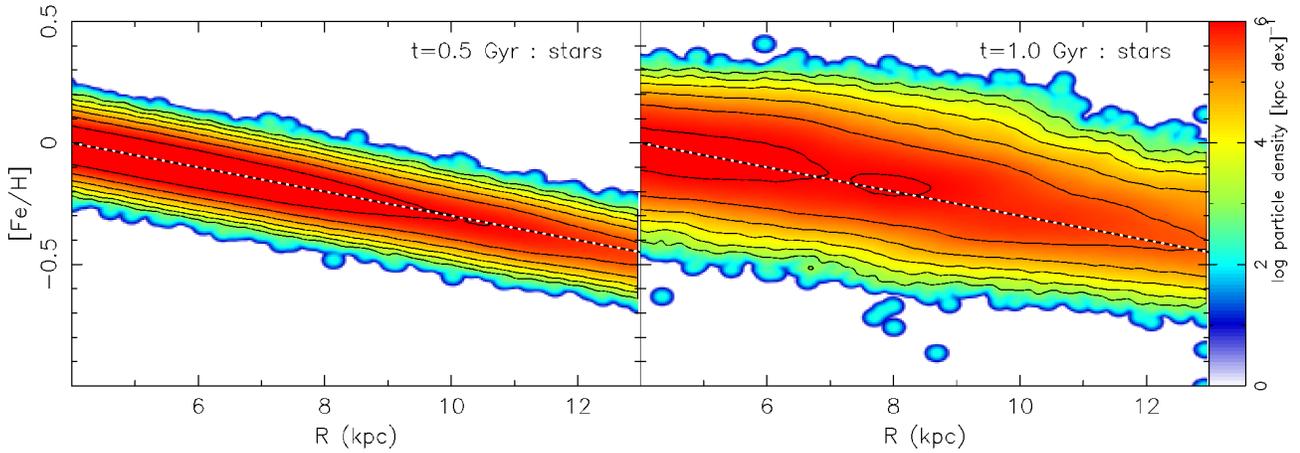} 
\caption{The same as Fig. \ref{fig2a} except that the metallicity of all star particles at $t=0.5$ Gyr is reset to the initial RMD set at $t=0$ Gyr (black dashed line). The radial metallicity gradient in both panels is very similar.}
\label{fig2ab}
\end{figure*}

A striking feature of radial migration around co-rotating spiral arms is that significant non-zero mean radial motion is evident at all radii. Fig. \ref{rthmap} shows the face-on gas density map and the mean velocity field in the $R-\theta$ plane at $t=0.7$ Gyr. Particularly clear around the spiral arm located between 20 and 100 degrees, the arrows indicate that the mean velocity of the gas behind the spiral arm is directed away from the galactic centre, over the whole range of the spiral arm. The opposite is true for gas located in front of the spiral arm. Qualitatively similar (though less strong) systematic motion is present in the stellar component, and is consistent with the large changes in angular momentum over all radii as shown in Fig. \ref{fig1}. This systematic streaming motion is a feature of co-rotating spiral arms that can be observationally tested by \emph{Gaia} \citep[e.g.][]{KHG14}.

Having established that radial migration occurs for both the star and gas particles, we now examine the effect of radial migration on the radial metallicity distribution (RMD) of both discs.

\begin{figure*}
\includegraphics[scale=0.12]{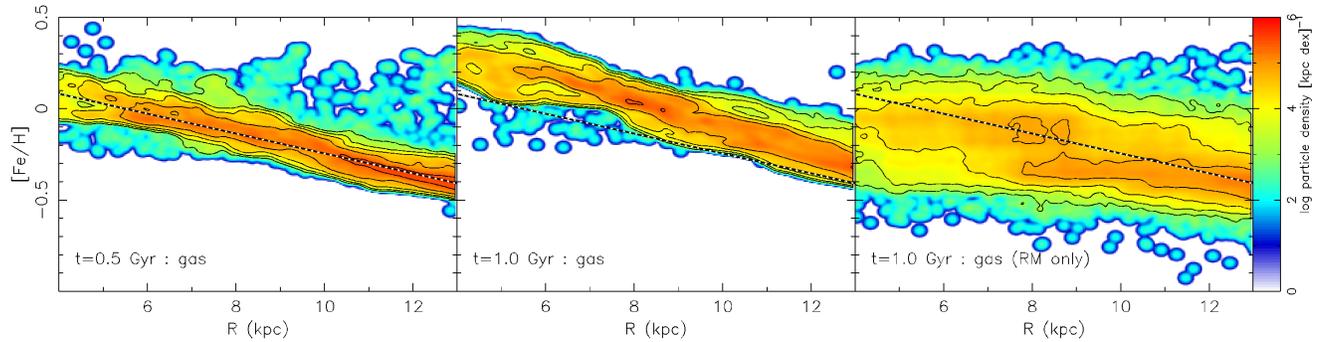}  
\caption{RMDs of gas particles that survive between $0.5$ and $1.0$ Gyr. \emph{Left:} The RMD at $t=0.5$ Gyr. \emph{Middle:} The RMD at $t=1.0$ Gyr. \emph{Right:} The radial position of gas particles at $t=1.0$ Gyr plotted with respect to their metallicity values at $t=0.5$ Gyr. This demonstrates the effect of radial migration only (i.e., no chemical evolution). The dashed line indicates the mean gas RMD at $t=0.5$ Gyr.}
\label{fig2b}
\end{figure*}

\subsection{Evolution of the stellar RMD}

To illustrate the effect of radial migration on the stellar RMD, we plot in Fig. \ref{fig2a} the RMD of all star particles at $t=0.5$ Gyr (left panel) and at $t=1.0$ Gyr (right panel). The distribution is broader at $t=1.0$ Gyr than at $t=0.5$ Gyr, although the broadening effect of radial migration is difficult to see because much of the broadening has already taken place in the first 0.5 Gyr of evolution when the bar forms. 

To illustrate more clearly the effect of radial migration on the RMD, at $t=0.5$ Gyr we reset the RMD to the initial RMD, i.e., the metallicity values of all star particles are re-assigned to a Gaussian distribution (of $0.05$ dex dispersion) at each radius with a mean metallicity value given by a metallicity gradient of $-0.05$ dex/kpc \citep[in the same way as shown in Fig. 13 of][]{GKC13b}. The resulting RMD is shown in the left panel of Fig. \ref{fig2ab}. The right panel of Fig. \ref{fig2ab} shows the RMD of all star particles at $t=1.0$ Gyr. The effects of radial migration on the RMD are now much more clear: the RMD shows clear broadening at all radii, while the radial metallicity gradient does not appear to change. This effect has been reported in other models presented in previous studies \citep[e.g.][]{SB02,ScB09,GKC13b}, and this confirms their findings.

\subsection{Evolution of the gas RMD}

In this section we examine the effects of radial migration and chemical evolution on the evolution of the gas RMD. The left panel of Fig. \ref{fig2b} shows the RMD of all gas particles at $t=0.5$ Gyr that are still gas particles at $t=1.0$ Gyr, which is clearly a narrow distribution around the mean metallicity at each radius (marked by the dashed line). The middle panel of Fig. \ref{fig2b} shows the RMD of the same gas particles at $t=1.0$ Gyr. At each radius the gas metallicity is higher than the mean metallicity of the gas at $t=0.5$ Gyr, which indicates that the gas metallicity has systematically increased over time. Because the simulation presented in this paper does not include infall of fresh metal poor gas from the inter-galactic medium onto the disc, the gas mass in the disc decreases with the time, and the metallicity of the gas increases artificially rapidly by metal enrichment. The systematic increase of the metallicity is artificial because of our numerical setup, and we cannot discuss the absolute value of the RMD in this numerical experiment. Therefore, our discussion below focuses on the relative difference between the gas and stellar RMD, and the dispersion of the gas RMD. 

\begin{figure}
\includegraphics[scale=0.43]{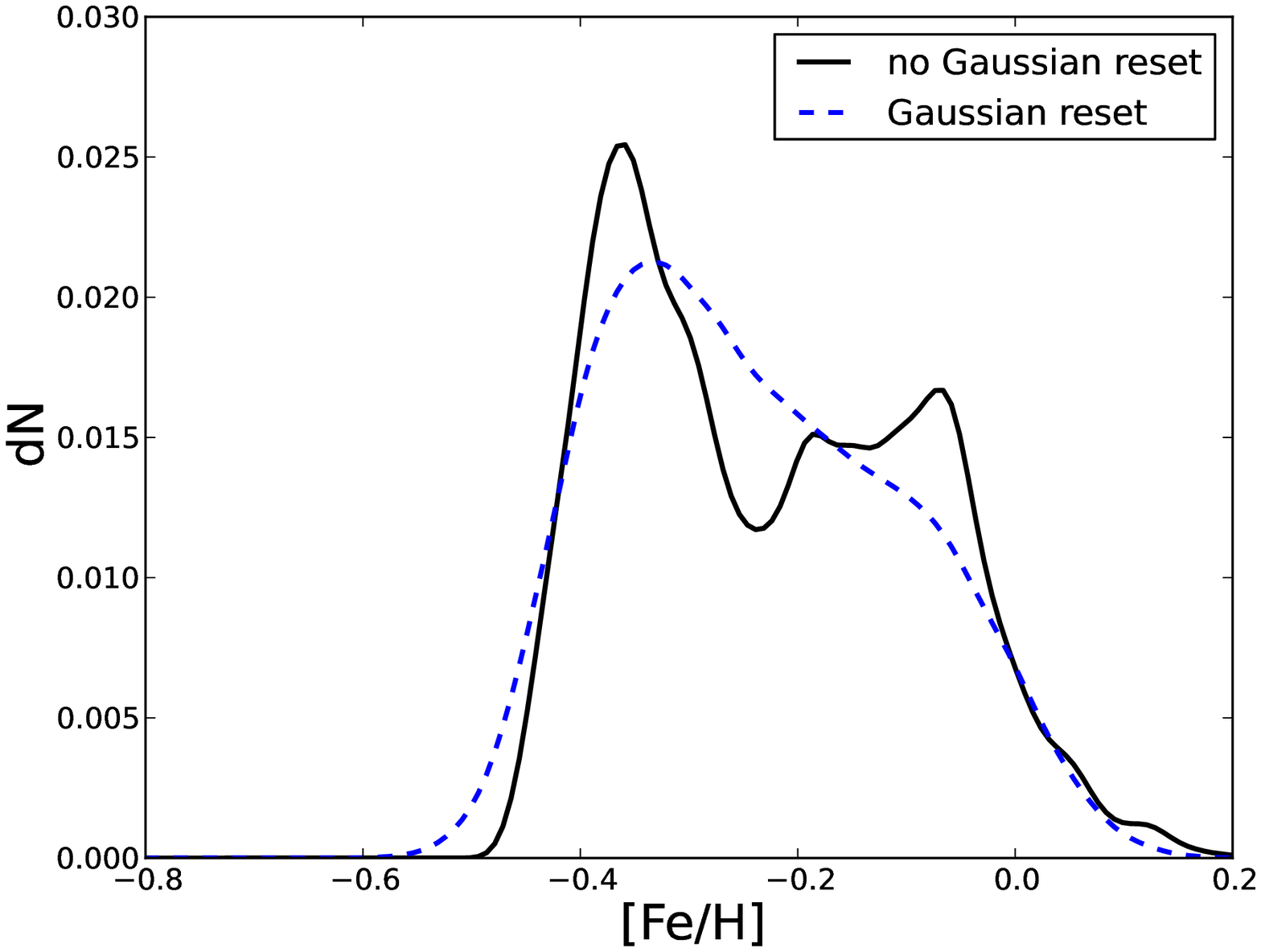}   
\caption{The metal distribution function of the gas within the radial range of $4<R<13$ kpc at $t=0.5$ Gyr (black solid line) and of the gas with metallicity distribution reset to follow the $t=0.5$ Gyr radial metallicity gradient of d[Fe/H]/dR$=-0.0545$ dex/kpc with a dispersion of $0.05$ dex (blue dashed line).}
\label{fig3}
\end{figure}

For comparison with the effect of radial migration only, the right panel of Fig. \ref{fig2b} shows the same gas particles at $t=1.0$ Gyr with the metallicity values they possessed at $t=0.5$ Gyr, i.e., chemical evolution is stopped for all gas particles after $t=0.5$ Gyr. In contrast to the combined effects of radial migration and chemical evolution shown in the middle panel, the RMD that results from radial migration alone is broad. However, the RMD shows a bimodal distribution in metallicity: one population around [Fe/H]$\sim-0.1$ and mainly distributed within $R=9$ kpc, and another population around [Fe/H]$=-0.4$ and mainly distributed outside $R=8$ kpc. The bimodal feature must be present at $t=0.5$ Gyr because the gas particles retain their respective $t=0.5$ Gyr metallicity values in the right panel of Fig. \ref{fig2b}. Although the gas at $t=0.5$ Gyr appears to follow only a single slope, the bimodal feature is revealed in Fig. \ref{fig3}, in which we show the metallicity distribution function (MDF) of stars that lie between $R=4$ and 13 kpc at $t=0.5$ Gyr (solid black curve). Subsequent radial migration of the gas particles stretches out the double peak feature in radius, which leads to the obvious bimodal feature in the $t=1.0$ Gyr RMD if we ignore the chemical evolution of the gas.

\begin{figure*}
\includegraphics[scale=0.16]{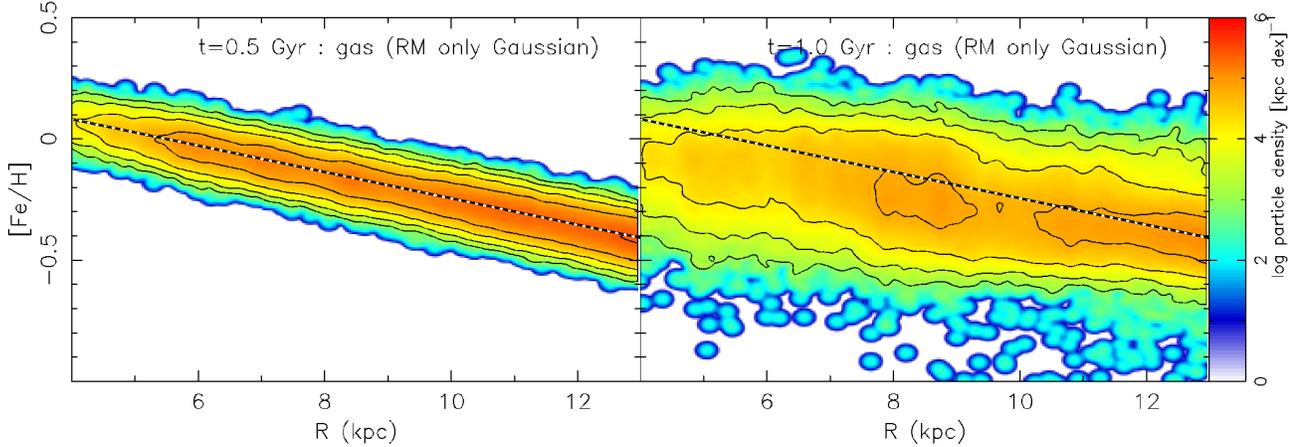}  
\caption{As for the left and right panels of Fig. \ref{fig2b} after the MDF at each radius is reset to a Gaussian distribution at $t=0.5$ Gyr. The dashed line indicates the mean gas RMD at $t=0.5$ Gyr.}
\label{fig2c}
\end{figure*}

The origin of the bimodal feature can be traced back to $t\sim 0.3$ Gyr when at bar is formed. The feature persists until $t\sim 0.7$ Gyr, after which times it disappears. Therefore we attribute the origin of the feature to the bar formation in the earlier epoch, and the bimodality at $t=0.5$ Gyr is therefore likely a relic of bar formation. During the bar formation, gas at radii inside $R\sim$10 kpc falls into the central regions of the disc \citep{MK13}, where chemical evolution likely proceeds differently from the outer region.

To focus on the effect of radial migration, we follow the procedure shown in Fig. ~\ref{fig2ab}: we reset the metallicity values of the gas particles such that, at each radius, the metallicity distribution function is set to a Gaussian distribtion with a dispersion of $0.05$ dex, and centred on the mean gas metallicity at $t=0.5$ Gyr. As shown by the dashed-blue line in Fig. \ref{fig3}, the MDF is smoothed and the bimodality is erased. The RMD at $t=0.5$ Gyr is shown in the left panel of Fig. \ref{fig2c}. The right panel of Fig. \ref{fig2c} shows the resulting gas RMD at $t=1.0$ Gyr, in which no bimodal feature is observed. As a result, the right panel of Fig.~\ref{fig2c} shows qualitatively similar evolution to the stars shown in the right panel of Fig. \ref{fig2ab}. This reinforces the broadening effect of radial migration on the metallicity distribution of the disc. The difference between the middle panel of Fig. \ref{fig2b} and the right panel of Fig.~\ref{fig2c} highlights that although gas radially migrates, the effect of chemical evolution acts to preserve the narrow {MDF} at all radii.

\subsection{The metallicity distribution function}
\label{sec3p1}

\begin{figure*}
\includegraphics[scale=0.16]{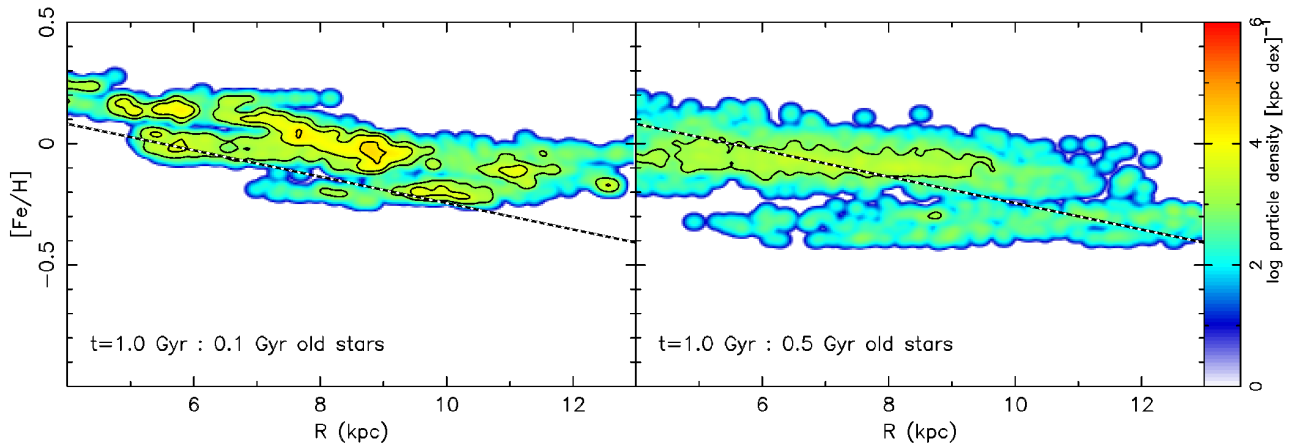} 
\caption{\emph{Left:} The RMD at $t=1.0$ Gyr of stars that were born between $0.9$ - $1.0$ Gyr. \emph{Right:} As for the left panel but for stars that were born between $0.45$ - $0.55$ Gyr. The younger stars trace the RMD of the gas in the middle panel of Fig. \ref{fig2b}, whereas the older stars trace the RMD of the gas in the right panel of Fig. \ref{fig2b} (see text for details). The dashed line indicates the mean gas RMD at $t=0.5$ Gyr.}
\label{figblah}
\end{figure*}

We now examine the consequences of the chemo-dynamical evolution of the stars and gas on the RMD of the stars with different ages. At $t=1.0$ Gyr, we select star particles born between 0.9 and 1.0 Gyr as a proxy for relatively younger stars, and star particles born between 0.45 and 0.55 Gyr as a proxy for a relatively older population of stars. We use the labels ``younger'' and ``older'' to describe these star particle populations to highlight their relative age difference, although both are young populations when considered with the stellar content of the Milky Way. The left panel of Fig. \ref{figblah} shows the RMD of the younger star particles at $t=1$ Gyr. Because the younger population of stars forms from the gas at around $t=1.0$ Gyr, the metallicity distribution of the younger star particles is narrow and traces the metallicity distribution of the gas at $t=1.0$ Gyr. This is qualitatively consistent with the observations reported in \citet{SBl09}. The RMD at $t=1.0$ Gyr of the older star particles is shown in the right panel of Fig. \ref{figblah}. The RMD of this relatively older population is broader than that of the younger population because once the star particles are formed between 0.45 and 0.55 Gyr their metal values do not change. The star particles are then subject to radial migration, which increases the metallicity dispersion at all radii in a similar way to the star particle RMD in Fig. \ref{fig2ab}. We note that the two apparently distinct groups of the older star particles, separated by the metallicity value $\rm[Fe/H]=-0.3$ dex, traces the gas metallicity distribution at $t=0.5$ Gyr. The two groups are therefore likely a relic of bar formation, as discussed in the previous section.

As noted above, our simulation does not include infall of fresh gas onto the disc, and the systematic increase in the gas metallicity seen in the left panel of Fig.~\ref{fig2b} is artificial. For example, the metal poor gas infall can lead to an RMD that does not change with time owing to the mixing of the metal poor infall gas with enriched disc gas, which is a more realistic scenario than the systematic increase in the gas metallicity. Therefore, we assume no systematic increase in the gas RMD, and compare the MDF of stars with different ages in this simulation by correcting the metallicity of the younger star particles. To apply the correction we subtract the difference in the mean metallicity of the gas at $t=0.5$ and 1.0 Gyr from the metallicity of the younger star particles.

\begin{figure}
\centering
\includegraphics[scale=0.44]{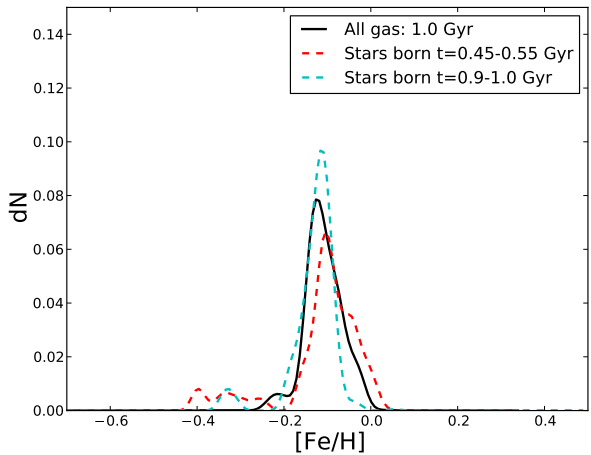}
\caption[]
{Metallicity distributions at the $R=8$ kpc radius of stars and gas populations at the times indicated in the legend. The MDF of the younger stars has been corrected according to the amount of chemical enrichment occurring in the gas since $t=0.5$ Gyr.}
\label{fig5}
\end{figure}

Fig. \ref{fig5} shows the MDF, at $t=1.0$ Gyr, of older star particles born between 0.45-0.55 Gyr and the enrichment-subtracted MDF of younger star particles born between 0.9-1.0 Gyr and all gas (with the same metallicity correction applied as the younger stars), at $R=8$ kpc. At $R=8$ kpc, the mean metallicity of the gas and younger star population at $t=1$ Gyr is higher than the mean gas metallicity at $t=0.5$ Gyr by 0.15 and 0.125 dex respectively. Therefore, we subtract 0.15 dex from the gas metallicity and 0.125 dex from the younger star metallicity at $t=1$ Gyr. The resulting MDF of the younger stars matches well that of the gas. The older star particle MDF (red) has a long metal-poor tail, and is broader than that of the younger star particles (cyan). Although we label these star populations as younger and older, both are young relative to the stellar content of the Milky Way. However, even within the $<1.0$ Gyr population of stars in the Milky Way there is an increase in the scatter of [Fe/H] as a function of age similar to what is observed in the solar neighbourhood \citep{CSA11}, which can be explained by the efficient radial migration in this paper.

\subsection{What keeps the metal dispersion so tight?}

\begin{figure*}
\includegraphics[scale=0.16]{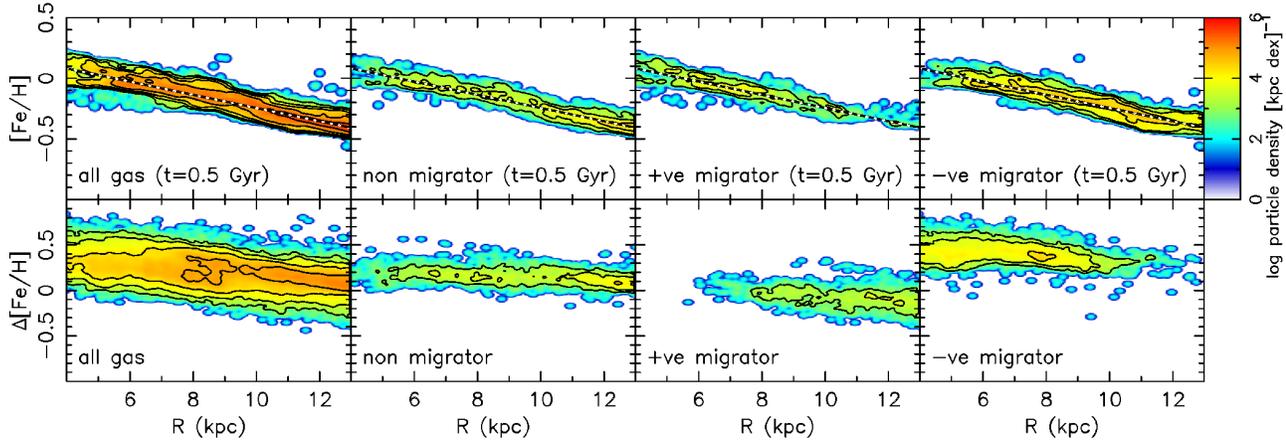} 
\caption{\emph{Top row:} The RMD of all gas particles (left), non-migrator gas particles (second), positive migrator gas particles (third) and negative migrator gas particles (right) at $t=0.5$ Gyr. The group of gas particles plotted is indicated in each panel. \emph{Bottom row:} The value of $\Delta$[Fe/H]$_i$ (described in the text) of the gas particle samples above, plotted as a function of particle radius at $t=1.0$ Gyr. Red colours indicate higher number density.}
\label{figdf}
\end{figure*}

We have shown above that despite the presence of radial migration in both the stars and gas, the evolution of the gas radial metallicity distribution is different from that of the stars. The gas metallicity shows a tighter relationship with the radius. At $t=1.0$ Gyr the gas RMD has a slightly steeper slope ($d{\rm [Fe/H]}/dR=-0.066$) than that at $t=0.5$ Gyr ($d{\rm [Fe/H]}/dR=-0.054$). To elucidate which process maintains the tightness of the gas RMD, we analyse the metal evolution of migrator and non-migrator gas particles that exist between $t=0.5$ to 1.0 Gyr. The non-migrator gas particles are defined as gas particles that have an angular momentum change, $\Delta L_z < 20.0$ kpc$^2$ Gyr$^{-1}$ between $t=0.5$ and 1.0 Gyr, and the positive ($+$ve) and negative ($-$ve) migrator particles are defined by $\Delta L_z > 830.0$ and $\Delta L_z < -830.0$ kpc$^2$ Gyr$^{-1}$, respectively. The RMDs of all gas particles, non-migrator gas particles and positive and negative migrator gas particles at $t=0.5$ Gyr are shown in the top row of Fig. \ref{figdf}. Of course, all particle samples follow the same narrow gas RMD with a negative radial metallicity gradient.

To quantify the metal evolution of each sample of particles, we define $\Delta$[Fe/H]$_i = $[Fe/H]$_i(t=1{\rm Gyr}) - $[Fe/H]$_i(t=0.5 {\rm Gyr})$ for each gas particle $i$ that exists between $t=0.5$ and 1.0 Gyr, and plot this quantity as a function of particle radius at $t=1.0$ Gyr for each particle sample in the second row of Fig. \ref{figdf}. The general positive $\Delta$[Fe/H]$_i$ of the total gas distribution shown in the bottom-left panel indicates that overall the gas is becoming more metal rich, which is consistent with the systematic increase in [Fe/H] seen in the middle panel of Fig.~\ref{fig2b}. The slight negative radial gradient of $\Delta$[Fe/H]$_i$ for all the gas particles is consistent with the slight steepening of the slope of the gas RMD at $t=1$ Gyr. Note that if there is no change of the slope in the RMD, the radial profile of $\Delta$[Fe/H]$_i$ should be flat.

The bottom-second panel of Fig. \ref{figdf} shows the metal change of non-migrator particles. Because these gas particles orbit around the same guiding centres during the period between 0.5 and 1.0 Gyr, the evolution of the RMD is mainly driven by chemical enrichment. The $\Delta$[Fe/H]$_i$ relation with radius for the non-migrator gas particles shows a tighter relation than that for all gas, and has a slightly negative slope, which is caused by more metal enrichment occurring in the inner disc. Note that without metal enrichment or radial migration, the effect of metal diffusion will be to flatten the gas radial metallicity gradient. This would lead to a positive slope in the $\Delta{\rm [Fe/H]}_i$ relation with radius.

The bottom-third and -fourth panels of Fig. \ref{figdf} show the $\Delta$[Fe/H]$_i$ for positive and negative migrator particles. The positive migrators show an overall decrease in metallicity in contrast to the chemically enriched non-migrator gas particles. The difference in the mean $\Delta$[Fe/H]$_i$ between the positive and non-migrator populations indicates that metal dilution has occurred in the positive migrator population. The dilution occurs because the positive migrator particles originate from the relatively more metal rich inner disc at $t=0.5$ Gyr (top-third panel of Fig.~\ref{figdf}). As they migrate to the outer radii where the gas is less metal rich, their metal content is diluted as it diffuses into nearby relatively metal poor gas in the outer disc.

Conversely, the negative migrators shown in the bottom-right panel show an increase in their metallicity larger than the amount that the non-migrator population has been enriched. The larger increase in metallicity compared to the non-migrator population is caused by the difference between their metallicity in the outer disc at $t=0.5$ Gyr (top-right panel of Fig.~\ref{figdf}) and the metallicity of the non-migrators in the inner disc at $t=1.0$ Gyr, in addition to metal mixing which is necessary in order for the metal distribution to maintain a dispersion. The metal mixing effect is approximately quantified by taking the difference between the mean $\Delta$[Fe/H]$_i$ value of the negative migrators and the non-migrators. 

At a given radius, we estimate the timescale of radial metal mixing caused by radial migration from the increase in the dispersion of the gas MDF between $t=0.5$ and $1.0$ Gyr, during which time all chemical evolution is stopped. As shown in Fig.~\ref{fig2c}, the metallicity distribution of the gas at $t=0.5$ Gyr is reset to follow the radial metallicity gradient with a Gaussian dispersion. We analyse the MDF evolution at $R=6$, $8$ and $10$ kpc. The radial mixing timescale owing to the radial migration is defined as the time taken for the dispersion of the MDF to increase by a factor of two. We find that the timescale is around $70$ Myr at all radii. \citet{YK12} suggested that thermal instability driven turbulent mixing can erase a metal inhomogeneity at scales of order $1$ kpc in a timescale of approximately $30$ Myr. Therefore, the local metal mixing in the gas from diffusion can mitigate the metal inhomogeneity induced by the radial migration of the gas as seen in our simulation.

It is therefore clear that chemical enrichment and metal mixing maintain the tight relation in the gas RMD by compensating the effects of radial migration that in general increase the scatter in the metallicity distribution at all radii. The mixing of metals provides a homogenising effect on the RMD: the metallicity dispersion is kept narrow at all radii. The more rapid metal enrichment in the inner region ensures the negative slope in the RMD.

\section{Conclusions}

We simulated the chemo-dynamical evolution of a stellar and gas disc of an isolated barred-spiral galaxy, similar in size to the Milky Way, and analysed the effect of radial migration on the metal distribution in both components. The main conclusions are as follows:

\begin{itemize}
\item The effect of stellar radial migration can be seen as a broadening of the metallicity distribution function at all radii. The RMD of all stars shows negligible change in metallicity gradient on the timescales studied in this paper.
\item Gas particles experience also radial migration. However, the MDF of the gas remains narrow at all radii. We examined the amount that each gas particle changed in metallicity between $0.5$ and 1.0 Gyr for a sample of positive, negative and non-migrator gas particles. The metal evolution of the non-migrator gas particles shows that the negative slope of the RMD is maintained by more efficient metal enrichment in the inner disc. The positive migrator gas particles generally decreased in metallicity. This is because these gas particles moved from the metal rich inner region into the metal poor outer region, therefore their metals are diluted into neighbouring particles. In contrast, gas particles that move from the metal poor outer regions into the metal rich inner regions (negative migrator gas particles) absorb metals from their neighbouring gas particles. Metal mixing causes the RMD to maintain a tight dispersion at all radii, whereas more metal enrichment in the inner region helps to keep the slope of the gas RMD negative.
\item We found that the width of the stellar MDF increases even in the short time scale of 0.5 Gyr, which is consistent with observations of solar neighbourhood stars \citep{CSA11}.
\end{itemize}

\section*{acknowledgements}
The authors thank the referee for a constructive report that led to improved clarity of the manuscript content. The calculations for this paper were performed on the UCL Legion, the Iridis HPC facility provided by the Centre for Innovation and the DiRAC Facilities (www.dirac.ac.uk, the DiRAC Shared Memory Processing system at the University of Cambridge operated by the COSMOS project, at the Department of Applied Mathematics and Theoretical Physics, the DiRAC Data Analytic system at the University of Cambridge, operated by the University of Cambridge High Performance Computing Service, the DiRAC Complexity system, operated by the University of Leicester IT Service, through the COSMOS consortium) jointly funded by BIS National E-infrastructure capital grant (ST/J0005673/1, ST/K001590/1 and ST/K000373/1), STFC capital grants (ST/H008861/1 and ST/H00887X/1) and STFC Operations grant (ST/K00333X/1). DiRAC is part of the National E-infrastructure. We also acknowledge PRACE for awarding us access to resource Cartesius based in Netherlands at SURFsara and Sisu based at CSC, Finland. This work was carried out, in part, through the Gaia Research for European Astronomy Training (GREAT-ITN) network. The research leading to these results has received funding from the European Union Seventh Framework Programme ([FP7/2007-2013] under grant agreement number 264895. 

\bibliographystyle{mn2e}
\bibliography{RMmetal-jan01.bbl}

\end{document}

%% file: aas_macros.tex
\def\jref@jnl#1{{\rm#1}}

\def\aj{\jref@jnl{AJ}}                   
\def\araa{\jref@jnl{ARA\&A}}             
\def\apj{\jref@jnl{ApJ}}                 
\def\apjl{\jref@jnl{ApJ}}                
\def\apjs{\jref@jnl{ApJS}}               
\def\ao{\jref@jnl{Appl.~Opt.}}           
\def\apss{\jref@jnl{Ap\&SS}}             
\def\aap{\jref@jnl{A\&A}}                
\def\aapr{\jref@jnl{A\&A~Rev.}}          
\def\aaps{\jref@jnl{A\&AS}}              
\def\azh{\jref@jnl{AZh}}                 
\def\baas{\jref@jnl{BAAS}}               
\def\jrasc{\jref@jnl{JRASC}}             
\def\memras{\jref@jnl{MmRAS}}            
\def\mnras{\jref@jnl{MNRAS}}             
\def\pasa{\jref@jnl{PASA}}
\def\pra{\jref@jnl{Phys.~Rev.~A}}        
\def\prb{\jref@jnl{Phys.~Rev.~B}}        
\def\prc{\jref@jnl{Phys.~Rev.~C}}        
\def\prd{\jref@jnl{Phys.~Rev.~D}}        
\def\pre{\jref@jnl{Phys.~Rev.~E}}        
\def\prl{\jref@jnl{Phys.~Rev.~Lett.}}    
\def\pasp{\jref@jnl{PASP}}               
\def\pasj{\jref@jnl{PASJ}}               
\def\qjras{\jref@jnl{QJRAS}}             
\def\skytel{\jref@jnl{S\&T}}             
\def\solphys{\jref@jnl{Sol.~Phys.}}      
\def\sovast{\jref@jnl{Soviet~Ast.}}      
\def\ssr{\jref@jnl{Space~Sci.~Rev.}}     
\def\zap{\jref@jnl{ZAp}}                 
\def\nat{\jref@jnl{Nature}}              
\def\iaucirc{\jref@jnl{IAU~Circ.}}       
\def\aplett{\jref@jnl{Astrophys.~Lett.}} 
\def\apspr{\jref@jnl{Astrophys.~Space~Phys.~Res.}}
\def\bain{\jref@jnl{Bull.~Astron.~Inst.~Netherlands}} 
\def\fcp{\jref@jnl{Fund.~Cosmic~Phys.}}  
\def\gca{\jref@jnl{Geochim.~Cosmochim.~Acta}}   
\def\grl{\jref@jnl{Geophys.~Res.~Lett.}} 
\def\jcp{\jref@jnl{J.~Chem.~Phys.}}      
\def\jgr{\jref@jnl{J.~Geophys.~Res.}}    
\def\jqsrt{\jref@jnl{J.~Quant.~Spec.~Radiat.~Transf.}}
\def\memsai{\jref@jnl{Mem.~Soc.~Astron.~Italiana}}
\def\nphysa{\jref@jnl{Nucl.~Phys.~A}}   
\def\physrep{\jref@jnl{Phys.~Rep.}}   
\def\physscr{\jref@jnl{Phys.~Scr}}   
\def\planss{\jref@jnl{Planet.~Space~Sci.}}   
\def\procspie{\jref@jnl{Proc.~SPIE}}   